\documentclass[aps,prl,twocolumn,superscriptaddress,showpacs,preprintnumbers,amsmath,amssymb]{revtex4-1}

\usepackage[dvipdfmx]{graphicx}
\usepackage{dcolumn}
\usepackage{subfigure}
\usepackage{color}
\usepackage[mathlines]{lineno}

\graphicspath{{ps}}

\begin{document}

\title{Measurement of the $\tau$ lepton polarization and $R(D^*)$\\in the decay $\bar{B} \to D^* \tau^- \bar{\nu}_\tau$}

\noaffiliation

\affiliation{University of the Basque Country UPV/EHU, 48080 Bilbao}
\affiliation{Beihang University, Beijing 100191}
\affiliation{University of Bonn, 53115 Bonn}
\affiliation{Budker Institute of Nuclear Physics SB RAS, Novosibirsk 630090}
\affiliation{Faculty of Mathematics and Physics, Charles University, 121 16 Prague}
\affiliation{Chonnam National University, Kwangju 660-701}
\affiliation{University of Cincinnati, Cincinnati, Ohio 45221}
\affiliation{Deutsches Elektronen--Synchrotron, 22607 Hamburg}
\affiliation{University of Florida, Gainesville, Florida 32611}
\affiliation{Justus-Liebig-Universit\"at Gie\ss{}en, 35392 Gie\ss{}en}
\affiliation{SOKENDAI (The Graduate University for Advanced Studies), Hayama 240-0193}
\affiliation{Hanyang University, Seoul 133-791}
\affiliation{University of Hawaii, Honolulu, Hawaii 96822}
\affiliation{High Energy Accelerator Research Organization (KEK), Tsukuba 305-0801}
\affiliation{J-PARC Branch, KEK Theory Center, High Energy Accelerator Research Organization (KEK), Tsukuba 305-0801}
\affiliation{IKERBASQUE, Basque Foundation for Science, 48013 Bilbao}
\affiliation{Indian Institute of Technology Bhubaneswar, Satya Nagar 751007}
\affiliation{Indian Institute of Technology Guwahati, Assam 781039}
\affiliation{Indian Institute of Technology Madras, Chennai 600036}
\affiliation{Indiana University, Bloomington, Indiana 47408}
\affiliation{Institute of High Energy Physics, Chinese Academy of Sciences, Beijing 100049}
\affiliation{Institute of High Energy Physics, Vienna 1050}
\affiliation{Institute for High Energy Physics, Protvino 142281}
\affiliation{INFN - Sezione di Torino, 10125 Torino}
\affiliation{J. Stefan Institute, 1000 Ljubljana}
\affiliation{Kanagawa University, Yokohama 221-8686}
\affiliation{Institut f\"ur Experimentelle Kernphysik, Karlsruher Institut f\"ur Technologie, 76131 Karlsruhe}
\affiliation{Kavli Institute for the Physics and Mathematics of the Universe (WPI), University of Tokyo, Kashiwa 277-8583}
\affiliation{Kennesaw State University, Kennesaw, Georgia 30144}
\affiliation{King Abdulaziz City for Science and Technology, Riyadh 11442}
\affiliation{Department of Physics, Faculty of Science, King Abdulaziz University, Jeddah 21589}
\affiliation{Korea Institute of Science and Technology Information, Daejeon 305-806}
\affiliation{Korea University, Seoul 136-713}
\affiliation{Kyoto University, Kyoto 606-8502}
\affiliation{Kyungpook National University, Daegu 702-701}
\affiliation{\'Ecole Polytechnique F\'ed\'erale de Lausanne (EPFL), Lausanne 1015}
\affiliation{P.N. Lebedev Physical Institute of the Russian Academy of Sciences, Moscow 119991}
\affiliation{Faculty of Mathematics and Physics, University of Ljubljana, 1000 Ljubljana}
\affiliation{Ludwig Maximilians University, 80539 Munich}
\affiliation{University of Maribor, 2000 Maribor}
\affiliation{Max-Planck-Institut f\"ur Physik, 80805 M\"unchen}
\affiliation{School of Physics, University of Melbourne, Victoria 3010}
\affiliation{Middle East Technical University, 06531 Ankara}
\affiliation{University of Miyazaki, Miyazaki 889-2192}
\affiliation{Moscow Physical Engineering Institute, Moscow 115409}
\affiliation{Moscow Institute of Physics and Technology, Moscow Region 141700}
\affiliation{Graduate School of Science, Nagoya University, Nagoya 464-8602}
\affiliation{Kobayashi-Maskawa Institute, Nagoya University, Nagoya 464-8602}
\affiliation{Nara Women's University, Nara 630-8506}
\affiliation{National Central University, Chung-li 32054}
\affiliation{National United University, Miao Li 36003}
\affiliation{Department of Physics, National Taiwan University, Taipei 10617}
\affiliation{H. Niewodniczanski Institute of Nuclear Physics, Krakow 31-342}
\affiliation{Nippon Dental University, Niigata 951-8580}
\affiliation{Niigata University, Niigata 950-2181}
\affiliation{Novosibirsk State University, Novosibirsk 630090}
\affiliation{Pacific Northwest National Laboratory, Richland, Washington 99352}
\affiliation{University of Pittsburgh, Pittsburgh, Pennsylvania 15260}
\affiliation{Punjab Agricultural University, Ludhiana 141004}
\affiliation{Theoretical Research Division, Nishina Center, RIKEN, Saitama 351-0198}
\affiliation{University of Science and Technology of China, Hefei 230026}
\affiliation{Showa Pharmaceutical University, Tokyo 194-8543}
\affiliation{Soongsil University, Seoul 156-743}
\affiliation{Stefan Meyer Institute for Subatomic Physics, Vienna 1090}
\affiliation{Sungkyunkwan University, Suwon 440-746}
\affiliation{School of Physics, University of Sydney, New South Wales 2006}
\affiliation{Department of Physics, Faculty of Science, University of Tabuk, Tabuk 71451}
\affiliation{Tata Institute of Fundamental Research, Mumbai 400005}
\affiliation{Excellence Cluster Universe, Technische Universit\"at M\"unchen, 85748 Garching}
\affiliation{Department of Physics, Technische Universit\"at M\"unchen, 85748 Garching}
\affiliation{Toho University, Funabashi 274-8510}
\affiliation{Department of Physics, Tohoku University, Sendai 980-8578}
\affiliation{Earthquake Research Institute, University of Tokyo, Tokyo 113-0032}
\affiliation{Department of Physics, University of Tokyo, Tokyo 113-0033}
\affiliation{Tokyo Institute of Technology, Tokyo 152-8550}
\affiliation{Tokyo Metropolitan University, Tokyo 192-0397}
\affiliation{University of Torino, 10124 Torino}
\affiliation{Virginia Polytechnic Institute and State University, Blacksburg, Virginia 24061}
\affiliation{Wayne State University, Detroit, Michigan 48202}
\affiliation{Yamagata University, Yamagata 990-8560}
\affiliation{Yonsei University, Seoul 120-749}
  \author{S.~Hirose}\affiliation{Graduate School of Science, Nagoya University, Nagoya 464-8602} 
  \author{T.~Iijima}\affiliation{Kobayashi-Maskawa Institute, Nagoya University, Nagoya 464-8602}\affiliation{Graduate School of Science, Nagoya University, Nagoya 464-8602} 
  \author{I.~Adachi}\affiliation{High Energy Accelerator Research Organization (KEK), Tsukuba 305-0801}\affiliation{SOKENDAI (The Graduate University for Advanced Studies), Hayama 240-0193} 
  \author{K.~Adamczyk}\affiliation{H. Niewodniczanski Institute of Nuclear Physics, Krakow 31-342} 
  \author{H.~Aihara}\affiliation{Department of Physics, University of Tokyo, Tokyo 113-0033} 
  \author{S.~Al~Said}\affiliation{Department of Physics, Faculty of Science, University of Tabuk, Tabuk 71451}\affiliation{Department of Physics, Faculty of Science, King Abdulaziz University, Jeddah 21589} 
  \author{D.~M.~Asner}\affiliation{Pacific Northwest National Laboratory, Richland, Washington 99352} 
  \author{H.~Atmacan}\affiliation{Middle East Technical University, 06531 Ankara} 
  \author{V.~Aulchenko}\affiliation{Budker Institute of Nuclear Physics SB RAS, Novosibirsk 630090}\affiliation{Novosibirsk State University, Novosibirsk 630090} 
  \author{T.~Aushev}\affiliation{Moscow Institute of Physics and Technology, Moscow Region 141700} 
% ...
\collaboration{The Belle Collaboration}

\noaffiliation


\begin{abstract}
  We report the first measurement of the $\tau$ lepton polarization $P_\tau(D^*)$ in the decay $\bar{B} \rightarrow D^* \tau^- \bar{\nu}_\tau$ as well as a new measurement of the ratio of the branching fractions $R(D^{*}) = \mathcal{B}(\bar {B} \rightarrow D^* \tau^- \bar{\nu}_\tau) / \mathcal{B}(\bar{B} \rightarrow D^* \ell^- \bar{\nu}_\ell)$, where $\ell^-$ denotes an electron or a muon, and the $\tau$ is reconstructed in the modes $\tau^- \rightarrow \pi^- \nu_\tau$ and $\tau^- \rightarrow \rho^- \nu_\tau$. We use the full data sample of $772 \times 10^6$ $B{\bar B}$ pairs recorded with the Belle detector at the KEKB electron-positron collider. Our results, $P_\tau(D^*) = -0.38 \pm 0.51 {\rm (stat.)} ^{+0.21} _{-0.16} {\rm (syst.)}$ and $R(D^*) = 0.270 \pm 0.035{\rm (stat.)} ^{+0.028}_{-0.025}{\rm (syst.)}$, are consistent with the theoretical predictions of the Standard Model.
\end{abstract}


\maketitle

\tighten

{\renewcommand{\thefootnote}{\fnsymbol{footnote}}}
\setcounter{footnote}{0}

Semileptonic $B$ decays to $\tau$ leptons (semitauonic decays) are theoretically well-studied processes within the Standard Model (SM)~\cite{cite:Heiliger:1989, cite:Korner:1990, cite:Hwang:2000}. The presence of the massive $\tau$ lepton in the decay increases the sensitivity to new physics (NP) beyond the SM, such as an extended Higgs sector. A prominent candidate is the Two-Higgs-Doublet Model (2HDM)~\cite{cite:2HDM:1989}, as suggested, for example, in Refs.~\cite{cite:Grzadkowski:1992,cite:Tanaka:1995,cite:Soni:1997,cite:Itoh:2005,cite:Crivellin:2012}, for the decay process $\bar{B} \to D^{(*)} \tau^- \bar{\nu}_\tau$~\cite{cite:CC}.

The decays $\bar{B} \to D^{(*)} \tau^- \bar{\nu}_\tau$ have been studied by the Belle~\cite{cite:Belle:2007, cite:Belle:2010, cite:Belle:2015, cite:Belle:2016}, BaBar~\cite{cite:BaBar:2008, cite:BaBar:2012:letter, cite:BaBar:2013:fullpaper} and LHCb~\cite{cite:LHCb:2015} experiments. Most of these studies have measured ratios of branching fractions, defined as $R(D^{(*)}) = \mathcal{B}(\bar{B} \to D^{(*)} \tau^- \bar{\nu}_\tau) / \mathcal{B}(\bar{B} \to D^{(*)} \ell^- \bar{\nu}_\ell)$. The denominator is the average of $\ell^- = e^-, \mu^-$ for Belle and BaBar, and $\ell^- = \mu^-$ for LHCb. The ratio cancels numerous uncertainties common to the numerator and the denominator. The current averages of the three experiments~\cite{cite:Belle:2015,cite:Belle:2016,cite:BaBar:2012:letter,cite:BaBar:2013:fullpaper,cite:LHCb:2015} are $R(D) = 0.397 \pm 0.040 \pm 0.028$ and $R(D^*) = 0.316 \pm 0.016 \pm 0.010$, which are $1.9$ and $3.3$ standard deviations ($\sigma$)~\cite{cite:HFAG:2014} away from the SM predictions of $R(D) = 0.299 \pm 0.011$~\cite{cite:RD_FermiandMILC:2015} or $0.300 \pm 0.008$~\cite{cite:RD_HPQCD:2015} and $R(D^*) = 0.252 \pm 0.003$~\cite{cite:RDst:2012}, respectively. The overall discrepancy with the SM is about 4$\sigma$. These tensions have been studied in the context of various NP models~\cite{cite:RDst:2012, cite:Datta:2012, cite:Tanaka:2013, cite:Biancofiore:2013, cite:Dorsner:2013, cite:Sakaki:2013, cite:Hagiwara:2014, cite:Duraisamy:2014, cite:Sakaki:2015, cite:Freytsis:2015, cite:Bhattacharya:2016, cite:Bardhan:2017}.

In addition to $R(D^{(*)})$, the polarizations of the $\tau$ lepton and the $D^*$ meson are also sensitive to NP~\cite{cite:Tanaka:1995, cite:Tanaka:2010, cite:RDst:2012, cite:Datta:2012, cite:Biancofiore:2013, cite:Tanaka:2013, cite:Sakaki:2013, cite:Duraisamy:2014, cite:Bhattacharya:2016, cite:Bardhan:2017}.  The $\tau$ lepton polarization is defined as $P_\tau(D^{(*)}) = [\Gamma^+(D^{(*)}) - \Gamma^-(D^{(*)})] / [\Gamma^+(D^{(*)}) + \Gamma^-(D^{(*)})]$, where $\Gamma^{\pm}(D^{(*)})$ denotes the decay rate of ${\bar B} \rightarrow D^{(*)} \tau^- {\bar \nu_{\tau}}$ with a $\tau$ helicity of $\pm 1/2$. The SM predicts $P_\tau(D) = 0.325 \pm 0.009$~\cite{cite:Tanaka:2010} and $P_\tau(D^*) = -0.497 \pm 0.013$~\cite{cite:Tanaka:2013}. For example, the type-II 2HDM allows $P_\tau(D^{(*)})$ to be between $-0.6$ ($-0.7$) and $+1.0$~\cite{cite:Tanaka:2013,cite:comment:2HDMII}. A leptoquark model suggested in Ref.~\cite{cite:Sakaki:2013} with a leptoquark mass of 1~TeV$/c^2$ is possible to take $P_\tau(D^*)$ between $-0.5$ and $0.0$. The $\tau$ polarization can be measured in two-body hadronic $\tau$ decays with the differential decay rate $[d \Gamma(D^{(*)}) / d \cos\theta_{\rm hel}] / \Gamma(D^{(*)}) = [1 + \alpha P_\tau(D^{(*)}) \cos \theta_{\rm hel}] / 2$, where $\theta_{\rm hel}$ is the angle of the $\tau$-daughter meson momentum with respect to the direction opposite the $W$ momentum in the rest frame of the $\tau$ (where $W$ denotes the $\tau^- \bar{\nu}_\tau$ system that corresponds to the virtual $W$ boson from the $B$ meson decay in the SM). The parameter $\alpha$ describes the sensitivity to $P_\tau(D^{(*)})$ for each $\tau$-decay mode; in particular, $\alpha = 1$ for $\tau^- \rightarrow \pi^- \nu_\tau$ and $\alpha = 0.45$ for $\tau^- \rightarrow \rho^- \nu_{\tau}$~\cite{cite:Hagiwara:1990}. In this Letter, we report the first $P_\tau(D^*)$ measurement in the decay $\bar{B} \rightarrow D^* \tau^- \bar{\nu}_\tau$ with the $\tau$ decays $\tau^- \rightarrow \pi^- \nu_{\tau}$ and $\tau^- \rightarrow \rho^- \nu_{\tau}$. Our study includes an $R(D^*)$ measurement independent of the previous studies~\cite{cite:Belle:2015, cite:Belle:2016, cite:BaBar:2012:letter, cite:BaBar:2013:fullpaper, cite:LHCb:2015}, in which leptonic $\tau$ decays have been used.

We use the full $\Upsilon(4S)$ data sample containing $772 \times 10^6 B\bar{B}$ pairs recorded with the Belle detector~\cite{cite:Belle-detector:2002} at the asymmetric-beam-energy $e^+ e^-$ collider KEKB~\cite{cite:KEKB:2003}. The Belle detector is a large-solid-angle magnetic spectrometer that consists of a silicon vertex detector (SVD), a 50-layer central drift chamber (CDC), an array of aerogel threshold Cherenkov counters (ACC), a barrel-like arrangement of time-of-flight scintillation counters (TOF), and an electromagnetic calorimeter (ECL) comprised of CsI(Tl) crystals located inside a superconducting solenoid coil that provides a 1.5~T magnetic field. An iron flux-return located outside of the coil is instrumented to detect $K_L^0$ mesons and to identify muons (KLM). The detector is described in detail elsewhere~\cite{cite:Belle-detector:2002}. 

The signal selection criteria are optimized using Monte Carlo (MC) simulation samples. These samples are generated using the software packages EvtGen~\cite{cite:EvtGen:2001} and PYTHIA~\cite{cite:PYTHIA:2006}, where final-state radiation is generated with PHOTOS~\cite{cite:PHOTOS:2016}. For the ${\bar B} \rightarrow D^* \tau^- {\bar\nu_{\tau}}$ (signal mode) and ${\bar B} \rightarrow D^* \ell^- {\bar\nu_{\ell}}$ (normalization mode) MC samples, we use hadronic form factors (FFs) based on heavy quark effective theory (HQET)~\cite{cite:Caprini:1998}. We use the world-average FF parameters extracted from $\bar{B} \rightarrow D^* \ell^- \bar{\nu}_\ell$ measurements~\cite{cite:HFAG:2014}. For the FF in a helicity-suppressed amplitude, which contributes negligibly in the light charged lepton mode, we adopt a theoretical estimate based on HQET~\cite{cite:RDst:2012}. Generated events are processed by the Belle detector simulator based on GEANT3~\cite{cite:GEANT:1984} to reproduce detector responses. 

We conduct the analysis by first identifying events where one of the two $B$ mesons ($B_{\rm tag}$) is reconstructed in one of 1104 exclusive hadronic $B$ decays using a hierarchical multivariate algorithm based on the NeuroBayes neural-network package~\cite{cite:Full-recon:2011}. More than 100 input variables are used to identify well-reconstructed $B$ candidates, including the difference $\Delta E \equiv E_{\rm tag}^* - E_{\rm beam}^*$ between the energy of the reconstructed $B_{\rm tag}$ candidate and the beam energy in the $e^+ e^-$ center-of-mass (CM) frame, as well as the event shape variables for suppression of $e^+ e^- \rightarrow q{\bar q}$ background ($q = u, d, s, c$). The quality of the $B_{\rm tag}$ candidate is synthesized in a single NeuroBayes output-variable classifier ($O_{\rm NB}$). We require the beam-energy-constrained mass of the $B_{\rm tag}$ candidate $M_{\rm bc} \equiv \sqrt{ E_{\rm beam}^{*2} / c^4 - |\vec{p}_{\rm tag}^{\kern2pt *}|^2 / c^2}$ (where $\vec{p}_{\rm tag}^{\kern2pt *}$ is the reconstructed $B_{\rm tag}$ three-momentum in the CM frame) to be greater than 5.272~GeV$/c^2$ and the value of $\Delta E$ to be between $-150$ and $100~{\rm MeV}$. We place a requirement on $O_{\rm NB}$ such that about 90\% of true $B_{\rm tag}$ and about 30\% of fake $B_{\rm tag}$ candidates are retained. If two or more $B_{\rm tag}$ candidates are retained in one event, we select the one with the highest $O_{\rm NB}$. The $B_{\rm tag}$ tagging efficiency is determined using the method described in Ref.~\cite{cite:Belle_Xulnu:2013}.

After $B_{\rm tag}$ selection, we form a signal-side $B$ candidate ($B_{\rm sig}$) from a $D^*$ candidate and a $\tau$ daughter or a charged-lepton candidate from the remaining particles. We use the following modes: $D^{*0} \rightarrow D^0 \gamma$, $D^0 \pi^0$, $D^{*+} \rightarrow D^+ \pi^0$ and $D^0 \pi^+$ for the $D^*$ candidate; $\tau^- \rightarrow \pi^- \nu_\tau$ and $\rho^- \nu_\tau$ for the $\tau$ candidate; $D^0 \rightarrow K_S^0 \pi^0$, $\pi^+ \pi^-$, $K^- \pi^+$, $K^+ K^-$, $K^- \pi^+ \pi^0$, $K_S^0 \pi^+ \pi^-$, $K_S^0 \pi^+ \pi^- \pi^0$, $K^- \pi^+ \pi^+ \pi^-$, $D^+ \rightarrow K_S^0 \pi^+$, $K_S^0 K^+$, $K_S^0 \pi^+ \pi^0$, $K^- \pi^+ \pi^+$, $K^+ K^- \pi^+$, $K^- \pi^+ \pi^+ \pi^0$ and $K_S^0 \pi^+ \pi^+ \pi^-$ for the $D$ candidate; and $K_S^0 \rightarrow \pi^+ \pi^-$, $\pi^0 \rightarrow \gamma \gamma$ and $\rho^- \rightarrow \pi^- \pi^0$, respectively, for the $K_S^0$, the $\pi^0$ and the $\rho$ meson candidates.

Charged particles are reconstructed using the SVD and the CDC; $K^\pm$, $\pi^\pm$ and $e^\pm$ candidates are identified based on the response of the inner detectors (CDC, TOF, ACC and ECL), while $\mu^\pm$ candidates are based on the responses in the CDC and the KLM. To form $K_S^0$ candidates~\cite{cite:Belle_KsKsKs:2005}, we combine pairs of oppositely-charged tracks with a vertex detached from the interaction point, impose pion mass hypotheses and require an invariant mass within $\pm$30~MeV/$c^2$ of the nominal $K_S^0$ mass~\cite{cite:PDG:2016}. Photons are reconstructed using ECL clusters not matched to charged tracks. Photon energy thresholds of 50, 100 and 150~MeV are used in the barrel, forward- and backward-endcap regions, respectively. Neutral pions are reconstructed from photon pairs with an invariant mass between 115 and 150 MeV/$c^2$. We impose tight selection criteria for $\pi^0$ from $D$ or $\rho$ (normal $\pi^0$) and looser criteria for $\pi^0$ from $D^*$ (soft $\pi^0$)~\cite{cite:Belle_Dlnu:2016}.

Candidate $D^{(*)}$ mesons are formed in the channels defined above. To maximize signal significance, $D$-mode-dependent invariant mass requirements are imposed. $D^*$ candidates are selected based on the $D^*$-mode-dependent mass difference $\Delta M \equiv M_{D^*} - M_D$ ($M_{D^{(*)}}$ being the invariant mass of the $D^{(*)}$ candidate).

For the $\pi^{\pm}$ candidates from $\tau$ decays, a proton veto is introduced to reduce baryonic peaking background such as ${\bar B} \rightarrow D^* {\bar p} n$ by about 80\% while retaining almost 100\% of the signal events. For the $\tau^- \to \rho^- \nu_{\tau}$ channel, $\rho$ candidates are formed from the combination of a $\pi^{\pm}$ and a $\pi^0$ with an invariant mass between 660 and 960~MeV$/c^2$. We then associate a $\pi^{\pm}$ or a $\rho^{\pm}$ candidate (one charged lepton) with the $D^*$ candidate to form signal (normalization) candidates. For the signal mode, the square of the momentum transfer $q^2 = (p_{e^+ e^-} - p_{\rm tag} - p_{D^*})^2$ (where $p$ denotes the four momentum) must be greater than 4~GeV$^2/c^2$. Finally, we require that there are no remaining charged tracks nor normal $\pi^0$ candidates in the event.

To measure $\cos \theta_{\rm hel}$, we first calculate the cosine of the angle between the momenta of the $\tau$ lepton and its daughter meson, $\cos\theta_{\tau d} = (2 E_\tau E_d - m_{\tau}^2 c^4 - m_d^2 c^4) / (2 |\vec{p}_{\tau}||\vec{p}_d| c^2)$ ($E$ and $\vec{p}$ being the energy and the three-momentum of the $\tau$ lepton or the $\tau$-daughter meson $d$), in the rest frame of the $\tau^- \bar{\nu}_\tau$ system. Using the Lorentz transformation from the rest frame of the $\tau^- \bar{\nu}_\tau$ system to the rest frame of $\tau$, the following equation is obtained: $|\vec{p}_d^{\kern2pt \tau}| \cos\theta_{\rm hel} = -\gamma |\vec{\beta}| E_d / c + \gamma |\vec{p}_d| \cos\theta_{\tau d}$, where $|\vec{p}_d^{\kern2pt \tau}| = (m_\tau^2 - m_d^2) / (2 m_\tau)$ is the $\tau$-daughter momentum in the rest frame of $\tau$, and $\gamma = E_\tau / (m_\tau c^2)$ and $|\vec{\beta}| = |\vec{p}_\tau| / E_\tau$. Solving this equation, the value of $\cos \theta_{\rm hel}$ is obtained. Events must lie in the physical region of $|\cos\theta_{\rm hel}| < 1$. To reject the $\bar{B} \rightarrow D^* \ell^- \bar{\nu}_\ell$ background in the $\tau^- \rightarrow \pi^- \nu_\tau$ sample, we only use the region $\cos\theta_{\rm hel} < 0.8$ in the fit.

After the event reconstruction, we find 1.03 to 1.09 candidates per event on average, depending on the signal mode. Most of the multiple-candidate events arise from more than one combination of a $D$ candidate with photons or soft pions. We select the best candidate based on the photon energy or the $\pi^0$ invariant mass in the $D^{*}$ candidate. Besides these, about 2\% of events are reconstructed both in the $\tau^- \rightarrow \pi^- \nu_{\tau}$ and $\rho^- \nu_{\tau}$ samples. Since the MC study indicates that 80\% of such events originate from the $\tau^- \rightarrow \rho^- \nu_\tau$ decay, we assign these events to the $\tau^- \rightarrow \rho^- \nu_{\tau}$ sample.

To separate signal events from background processes, we use the variable $E_{\rm ECL}$, the linearly-summed energy of ECL clusters not used in the reconstruction of the $B_{\rm sig}$ and $B_{\rm tag}$ candidates. For normalization events with charged lepton $\ell$, we use the variable $M_{\rm miss}^2 = (p_{e^+ e^-} - p_{\rm tag} - p_{D^*} - p_{\ell})^2 / c^2$ as its values populate the region near $M_{\rm miss}^2 = 0$. We use the MC distributions of these variables as the histogram probability density functions (PDFs) in the final fit. The signal PDF is validated using the normalization sample. We find good agreement between the data and the MC distributions for $E_{\rm ECL}$. The $M_{\rm miss}^2$ resolution in the data is slightly worse than in the MC. We therefore broaden the width of the peaking component in the $M_{\rm miss}^2$ signal PDF to match that of the data.

The most significant irreducible background contribution is from events with incorrectly-reconstructed $D^*$ candidates, denoted ``fake $D^*$.'' We compare the PDF shapes of these events in $\Delta M$ sideband regions. While we find good agreement of the $E_{\rm ECL}$ shapes between the data and the MC, we observe a slight discrepancy in the $M_{\rm miss}^2$ shape. The $M_{\rm miss}^2$ discrepancy is corrected based on this comparison.

Semileptonic decays to excited charm modes, ${\bar B} \rightarrow D^{**} \ell^- \bar{\nu}_{\ell}$ and ${\bar B} \rightarrow D^{**} \tau^- \bar{\nu}_\tau$, generally represent an important background in the $\bar{B} \rightarrow D^* \tau^- \bar{\nu}_\tau$ study as they have a similar decay topology to the signal events. Moreover, background events from various types of hadronic $B$ decays wherein some particles are not reconstructed are significant in our measurement. Since there are many unmeasured exclusive modes of these $B$ decays and hence a large uncertainty in the yield, we determine their yields in the final fit to data. The PDF shape uncertainty of these backgrounds is taken into account, as a change in the $B$ decay composition may modify the $E_{\rm ECL}$ shape and thereby introduce biases in the measurement of $R(D^*)$ and $P_\tau(D^*)$. For the decays with experimentally-measured branching fractions, we use the values in Refs.~\cite{cite:PDG:2016, cite:Belle_DstKKL:2002, cite:Belle_Dstomegapi:2015}. Other types of hadronic $B$ decay background often contain neutral particles such as $\pi^0$ and $\eta$ or pairs of charged pions. We calibrate the composition of hadronic $B$ decays in the MC based on calibration data samples by reconstructing seven final states (${\bar B} \rightarrow D^* \pi^- \pi^- \pi^+$, $D^* \pi^- \pi^- \pi^+ \pi^0$, $D^* \pi^- \pi^- \pi^+ \pi^0 \pi^0$, $D^* \pi^- \pi^0$, $D^* \pi^- \pi^0 \pi^0$, $D^* \pi^- \eta$, and $D^* \pi^- \eta \pi^0$) in the signal-side. Candidate $\eta$ mesons are reconstructed using pairs of photons with an invariant mass ranging from 500 to 600~MeV$/c^2$. We then extract the calibration sample yield with the signal-side energy difference $\Delta E^{\rm sig}$ or the beam-energy-constrained mass $M_{\rm bc}^{\rm sig}$ in the region $q^2 > 4~{\rm GeV}^2/c^2$ and $|\cos\theta_{\rm hel}| < 1$. To calculate $\cos \theta_{\rm hel}$, we assume that (one of) the charged pion(s) is the $\tau$ daughter. We use a ratio of the yield in the data to that in the MC as the yield scale factor. If there is no observed event in the calibration sample, we assign a 68\% confidence level upper limit on the scale factor. The above calibrations cover about 80\% of the hadronic $B$ background. For the remaining $B$ decay modes, we assume 100\% uncertainty on the MC expectation.

In the signal extraction, we consider three ${\bar B} \rightarrow D^* \tau^- {\bar \nu_{\tau}}$ components: (i) the ``signal'' component contains correctly-reconstructed signal events, (ii) the ``$\rho \leftrightarrow \pi$ cross feed'' component contains events where the decay $\tau^- \rightarrow \rho^- (\pi^-) \nu_{\tau}$ is reconstructed as $\tau^- \rightarrow \pi^- (\rho^-) \nu_{\tau}$, (iii) the ``other $\tau$ cross feed'' component contains events with other $\tau$ decays such as $\tau^- \rightarrow \mu^- {\bar \nu_{\mu}} \nu_{\tau}$ and $\tau^- \rightarrow \pi^- \pi^0 \pi^0 \nu_{\tau}$. The relative contributions are fixed based on the MC. We relate the signal yield and $R(D^*)$ as $R(D^*) = (\epsilon_{\rm norm}N_{\rm sig}) / (\mathcal{B}_\tau\epsilon_{\rm sig}N_{\rm norm})$, where $\mathcal{B}_{\tau}$ denotes the branching fraction of $\tau^- \rightarrow \pi^- \nu_{\tau}$ or $\tau^- \rightarrow \rho^- \nu_{\tau}$, and $\epsilon_{\rm sig}$ and $\epsilon_{\rm norm}$ ($N_{\rm sig}$ and $N_{\rm norm}$) are the efficiencies (the observed yields) for the signal and the normalization mode. Using the MC, the efficiency ratio $\epsilon_{\rm norm} / \epsilon_{\rm sig}$ of the signal component in the $B^-$ ($\bar{B}^0$) sample is estimated to be $0.97 \pm 0.02$ ($1.21 \pm 0.03$) for the $\tau^- \rightarrow \pi^- \nu_\tau$ mode and $3.42 \pm 0.07$ ($3.83 \pm 0.12$) for the $\tau^- \rightarrow \rho^- \nu_\tau$ mode, where the quoted errors arise from MC statistical uncertainties. The larger efficiency ratio for the $\bar{B}^0$ mode is due to the significant $q^2$ dependence of the efficiency in the $D^{*+} \rightarrow D^0 \pi^+$ mode. For $P_\tau(D^*)$, we divide the signal sample into two regions $\cos\theta_{\rm hel} > 0$ (forward) and $\cos\theta_{\rm hel} < 0$ (backward). The value of $P_\tau(D^*)$ is then parameterized as $P_\tau(D^*) = [2 (N_{\rm sig}^{\rm F} - N_{\rm sig}^{\rm B})] / [\alpha (N_{\rm sig}^{\rm F} + N_{\rm sig}^{\rm B})]$, where the superscript F (B) denotes the signal yield in the forward (backward) region. The detector bias on $P_\tau(D^*)$ is taken into account with a linear function that relates the true $P_\tau(D^*)$ to the extracted $P_\tau(D^*)$ ($P_\tau(D^*)$ correction function), determined using several MC sets with different $P_\tau(D^*)$ values. Here, other kinematic distributions are assumed to be consistent with the SM prediction.

\begin{figure}[t!]
  \centering
  \includegraphics[angle=90,width=6cm]{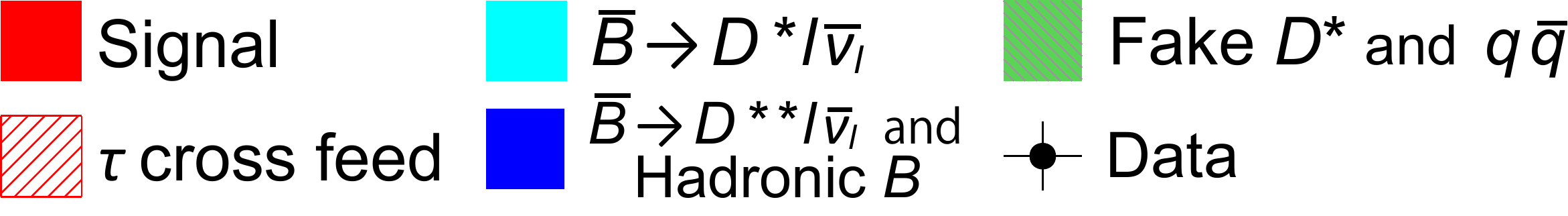}
  \includegraphics[width=7cm]{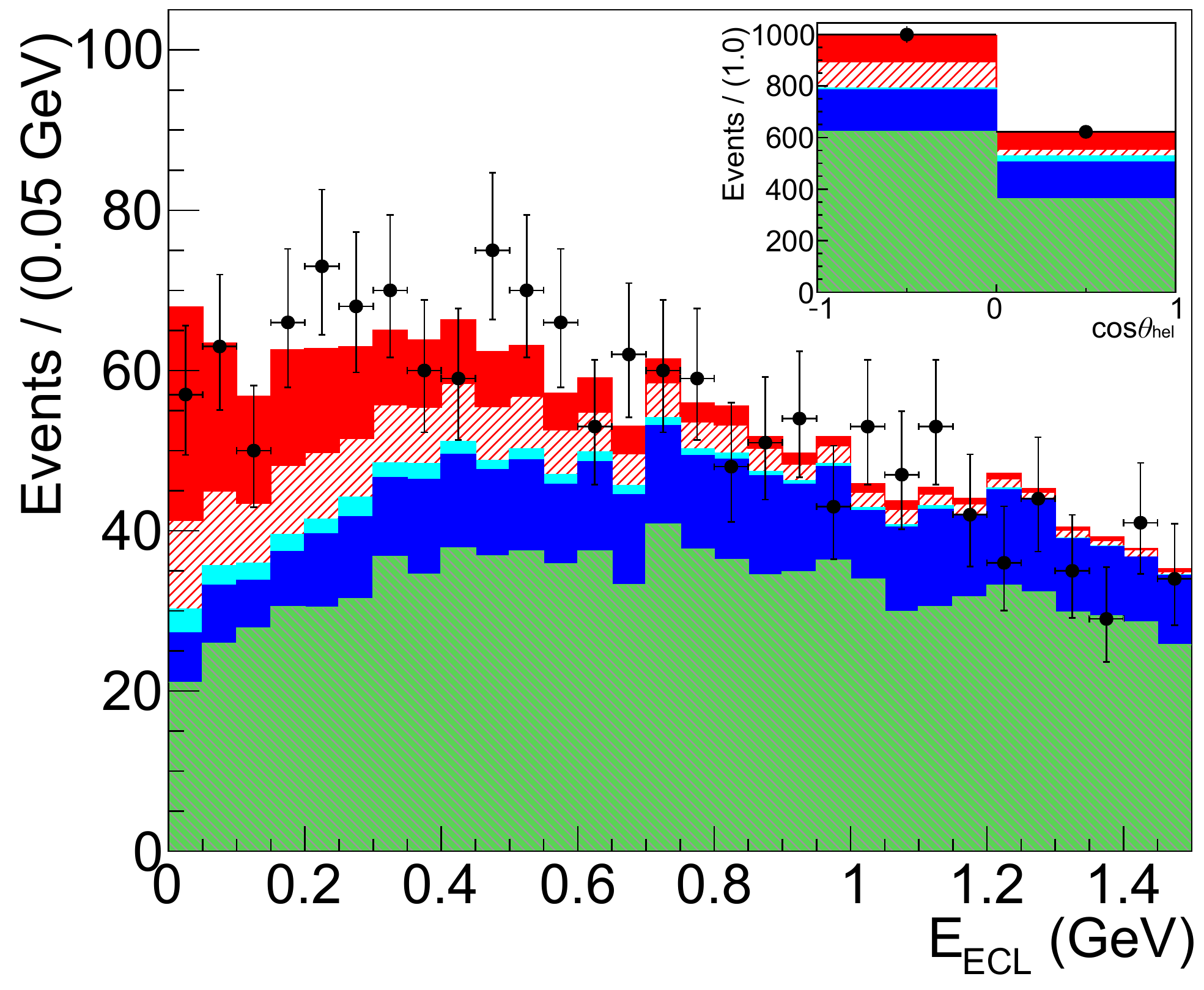}
  \caption{Fit result to the signal sample (all the eight samples are combined). The main panel and the sub panel show the $E_{\rm ECL}$ and the $\cos\theta_{\rm hel}$ distributions, respectively. The red-hatched ``$\tau$ cross feed'' combines the $\rho \leftrightarrow \pi$ cross-feed and the other $\tau$ cross-feed components.}
  \label{fig:fit-result}
\end{figure}

We categorize the background into four components. The ``${\bar B} \rightarrow D^* \ell^- {\bar \nu_{\ell}}$'' component contaminates the signal sample due to the misassignment of the lepton as a pion. We fix the ${\bar B} \rightarrow D^* \ell^- {\bar \nu_{\ell}}$ background yield from the fit to the normalization sample. For the ``${\bar B} \rightarrow D^{**} \ell^- {\bar \nu_{\ell}}$ and hadronic $B$ decay'' component, we combine all the modes into common yield parameters. One exception is the decay into two $D$ mesons such as ${\bar B} \rightarrow D^* D_s^{*-}$ and ${\bar B} \rightarrow D^* {\bar D^{(*)}} K^-$. Since these decays are experimentally well measured, we fix their yields based on the world-average branching fractions~\cite{cite:PDG:2016}. The yield of the ``fake $D^*$'' component is fixed from a comparison of the data and the MC in the $\Delta M$ sideband regions. The contribution from the continuum $e^+ e^- \rightarrow q{\bar q}$ process is only $\mathcal{O}(0.1\%)$. We therefore fix the yield using the MC expectation. 

We then conduct an extended binned maximum likelihood fit in two steps; we first perform a fit to the normalization sample to determine its yield, and then a simultaneous fit to eight signal samples $(B^-, \bar{B}^0) \otimes (\pi^- \nu_\tau, \rho^- \nu_\tau) \otimes ({\rm backward, forward})$. In the fit, $R(D^*)$ and $P_\tau(D^*)$ are common fit parameters, while the ``$\bar{B} \rightarrow D^{**} \ell^- \bar{\nu}_\ell$ and hadronic $B$'' yields are independent among the eight signal samples. The fit result is shown in Fig.~\ref{fig:fit-result}. The obtained signal and normalization yields for $B^-$ ($\bar{B}^0$) mode are, respectively, $210 \pm 27$ ($88 \pm 11$) and $4711 \pm 81$ ($2502 \pm 52$), where the errors are statistical.

\begin{figure}[t!]
  \centering
  \includegraphics[width=8cm]{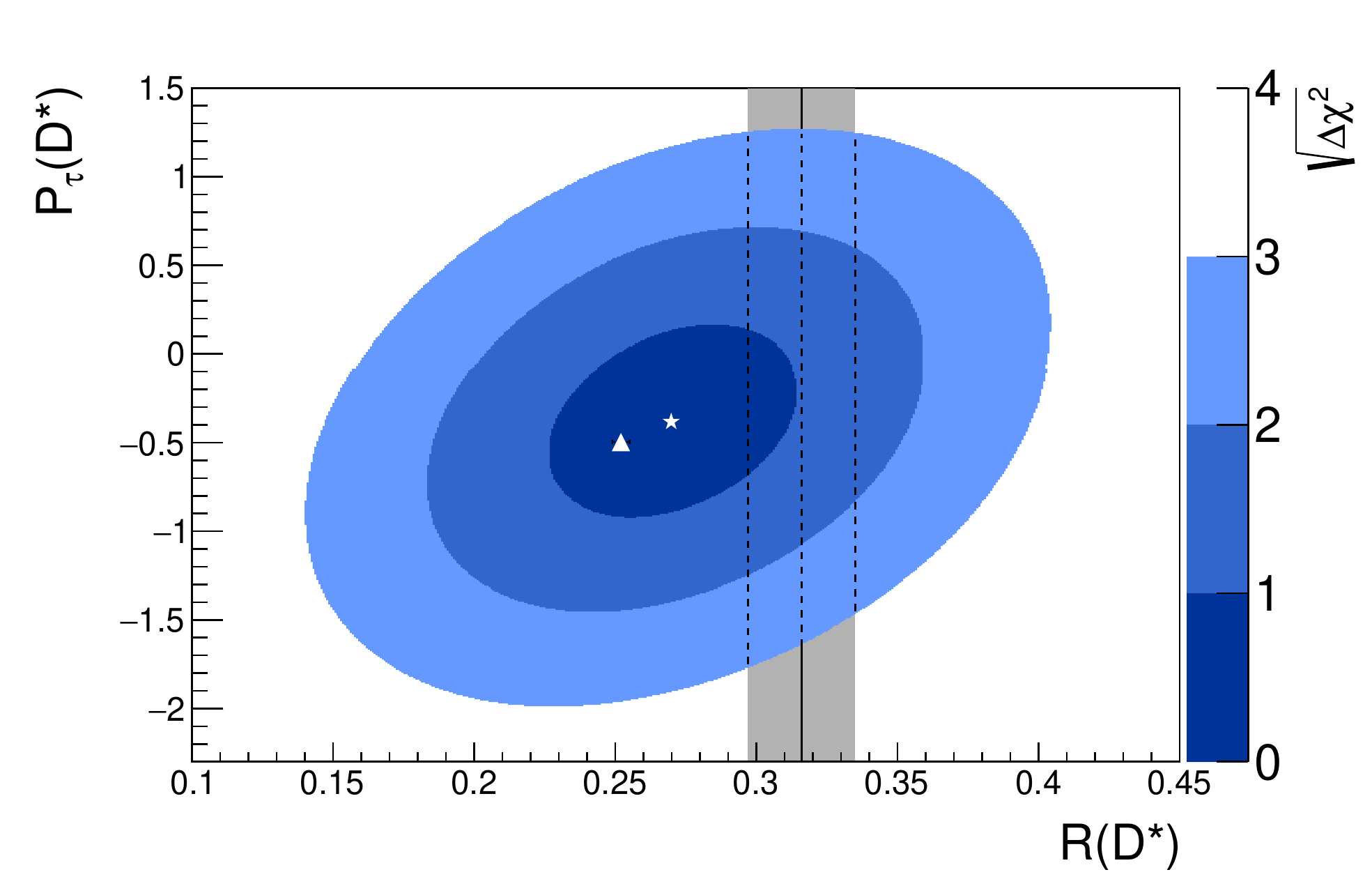}
  \caption{Comparison of our result (star for the best-fit value and 1$\sigma$, 2$\sigma$, 3$\sigma$ contours) with the SM prediction~\cite{cite:RDst:2012, cite:Tanaka:2013} (triangle). The shaded vertical band shows the world average~\cite{cite:HFAG:2014} without our result.}
  \label{fig:significance}
\end{figure}

The most significant systematic uncertainty arises from the hadronic $B$ decay composition ($^{+7.7}_{-6.9}\%,\,{}^{+0.13}_{-0.10}$), where the first (second) value in the parentheses is the relative (absolute) uncertainty in $R(D^*)$ ($P_\tau(D^*)$). The limited MC sample size used in the analysis introduces statistical fluctuations on the PDF shapes $(^{+4.0}_{-2.8}\%,\,{}^{+0.15}_{-0.11})$. The uncertainties arising from the semileptonic $B$ decays are $(\pm 3.5\%, \pm 0.05)$. The fake $D^*$ background, which dominates in this analysis, causes uncertainties of ($\pm 3.4\%, \pm 0.02$). Other uncertainties arise from the reconstruction efficiencies for the $\tau$ daughter and the charged lepton, the signal and normalization efficiencies, the choice of the number of bins in the fit, the $\tau$ branching fractions and the $P_\tau(D^*)$ correction function parameters. These systematic uncertainties account for ($\pm 2.2\%, \pm 0.03$). In addition, since we fix part of the background yield, we need to consider the impact from the uncertainties that are common between the signal and the normalization: the number of $B \bar{B}$ events, the tagging efficiency, the $D$ branching fractions and the $D^*$ reconstruction efficiency. The total for this source is ($\pm 2.3\%, \pm 0.02$). In the calculation of the total systematic uncertainty, we treat the systematic uncertainties as independent, except for those of the $\tau$ daughter and the $D^*$ reconstruction efficiencies. The latter originate from the same sources: the particle-identification efficiencies for $K^\pm$ and $\pi^\pm$ and the reconstruction efficiencies for $K_S^0$ and $\pi^0$. We therefore account for this correlation. The total systematic uncertainties are ($^{+10.4}_{-9.4}\%,\,{}^{+0.21}_{-0.16}$). The final results, shown in Fig.~\ref{fig:significance}, are:
\begin{eqnarray}
  R(D^*)   &=& 0.270 \pm 0.035 ({\rm stat.}) ^{+0.028} _{-0.025} ({\rm syst.}),\nonumber\\
  P_\tau(D^*) &=& -0.38 \pm 0.51 ({\rm stat.}) ^{+0.21} _{-0.16} ({\rm syst.}).\nonumber
\end{eqnarray}
The statistical correlation is 0.29, and the total correlation (including systematics) is 0.33. Overall, our result is consistent with the SM prediction. The obtained $R(D^*)$ is independent of and also agrees with the previous Belle measurements, $R(D^*) = 0.293 \pm 0.038 \pm 0.015$~\cite{cite:Belle:2015} and $0.302 \pm 0.030 \pm 0.011$~\cite{cite:Belle:2016}, and with the world average~\cite{cite:HFAG:2014}. Moreover, our measurement excludes $P_\tau(D*) > +0.5$ at 90\% C.L.

In summary, we report a measurement of $P_\tau(D^*)$ in the decay ${\bar B} \rightarrow D^* \tau^- {\bar \nu_{\tau}}$ as well as a new $R(D^*)$ measurement with the hadronic $\tau$ decay modes $\tau^- \rightarrow \pi^- \nu_{\tau}$ and $\tau^- \rightarrow \rho^- \nu_{\tau}$, using $772 \times 10^6$ $B\bar{B}$ events recorded with the Belle detector. Our results, $R(D^*) = 0.270 \pm 0.035{\rm (stat.)}\ {}^{+0.028}_{-0.025}{\rm (syst.)}$ and $P_\tau(D^*) = -0.38 \pm 0.51 {\rm (stat.)} \ {}^{+0.21} _{-0.16} {\rm (syst.)}$, are consistent with the SM prediction. We have measured $P_\tau(D^*)$ for the first time, which provides a new dimension in the search for NP in semitauonic $B$ decays.

We acknowledge Y.~Sakaki, M.~Tanaka and R.~Watanabe for their invaluable suggestions and help.

We thank the KEKB group for excellent operation of the
accelerator; the KEK cryogenics group for efficient solenoid
operations; and the KEK computer group, the NII, and
PNNL/EMSL for valuable computing and SINET5 network support.
We acknowledge support from MEXT, JSPS and Nagoya's TLPRC (Japan);
ARC (Australia); FWF (Austria); NSFC and CCEPP (China);
MSMT (Czechia); CZF, DFG, EXC153, and VS (Germany);
DST (India); INFN (Italy);
MOE, MSIP, NRF, BK21Plus, WCU, RSRI, FLRFAS project and GSDC of KISTI (Korea);
MNiSW and NCN (Poland); MES and RFAAE (Russia); ARRS (Slovenia);
IKERBASQUE and UPV/EHU (Spain);
SNSF (Switzerland); MOE and MOST (Taiwan); and DOE and NSF (USA).

This work is supported by a Grant-in-Aid for Scientific Research (S) ``Probing New Physics with Tau-Lepton'' (No. 26220706) 
and was partly supported by a Grant-in-Aid for JSPS Fellows (No. 25.3096).


\begin{thebibliography}{99}
\bibitem{cite:CC}
  Throughout this Letter, the inclusion of the charge-conjugate mode is implied.
\end{thebibliography}
\end{document}